\documentclass{WileyMSP-template}

\usepackage{xr-hyper}
\usepackage{color}

\makeatletter
\newcommand*{\addFileDependency}[1]{
	\typeout{(#1)}
	\@addtofilelist{#1}
	\IfFileExists{#1}{}{\typeout{No file #1.}}
}
\makeatother
\newcommand*{\myexternaldocument}[1]{%
	\externaldocument{#1}%
	\addFileDependency{#1.tex}%
	\addFileDependency{#1.aux}%
}

\graphicspath{X:\paper
}
\myexternaldocument{SupplementaryInformation}

\begin{document}

\pagestyle{fancy}
\rhead{\includegraphics[width=2.5cm]{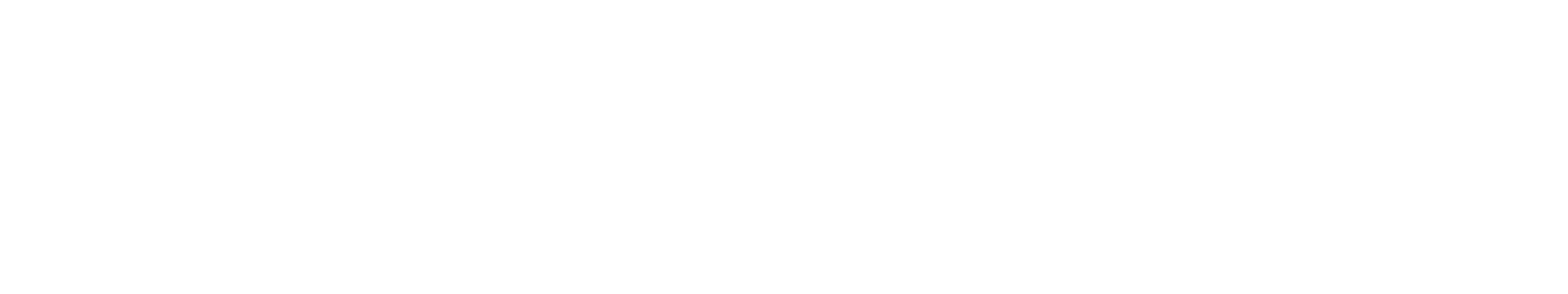}}

\title{Probing spatial variation of magnetic order in strained SrMnO$_3$ thin films using Spin Hall Magnetoresistance}

\maketitle

\author{J. J. L. van Rijn*}
\author{T. Banerjee*}

\begin{affiliations}
J.J.L. van Rijn, Prof. T. Banerjee\\
University of Groningen, Zernike Institute for Advanced Materials, 9747 AG Groningen, The Netherlands\\
Email Address: j.j.l.van.rijn@rug.nl, t.banerjee@rug.nl

\end{affiliations}

\keywords{Complex oxides, Antiferromagnets, Multiferroic, Spin Hall magnetoresistance}

\begin{abstract}

SrMnO$_{3}$ (SMO) is a magnetic insulator and predicted to exhibit a multiferroic phase upon straining. Strained films of SMO display a wide range of magnetic orders, ranging from G-type to C-and A-type, indicative of competing magnetic interactions. The potential of spin Hall magnetoresistance (SMR) is exploited as an electrical probe for detecting surface magnetic order, to read surface magnetic moments in SMO and its spatial variation, by designing and positioning electrodes of different sizes on the film. The findings demonstrate antiferromagnetic domains with different magnetocrystalline anisotropies along with a ferromagnetic order, where the magnetization arises from double exchange mediated ferromagnetic order and canted antiferromagnetic moments. Further, from a complete analysis of the SMR, a predominance of antiferromagnetic domain sizes of 3.5 $\mu$m$^2$ is extracted. This work enhances the applicability of SMR in unraveling the richness of correlation effects in complex oxides, as manifested by the detection of coexisting and competing ground states and lays the foundation for the study of magnon transport for different magnetoelectric based computing applications.

\end{abstract}

\section{Introduction}

The emergence of coexisting  magnetic orders that couples with other ferroic orders is a thriving playground for the exploration of topological textures in materials. Multiferroic materials, due to their intrinsic magnetoelectric coupling are natural platforms for this and serve as building blocks for low power computing applications \cite{manipatruni2019scalable,scott2007multiferroic}. Strain control in thin films is an established method that enriches and enhances this coupling and has recently invigorated research in this direction \cite{spaldin2019advances,gupta2022review}.  

Complex oxide materials are a versatile material class where strong correlation between structural, magnetic and electronic properties can be exploited to create emergent phases that can be tailored by varying the strain quotient. Our material of choice, SrMnO$_3$ (SMO), is an emerging magnetic insulator in this class and predicted to exhibit multiferroicity upon straining \cite{lee2010epitaxial,edstrom2018first,edstrom2020prediction,zhu2020magnetic}. Interestingly, significant magnetoelectric coupling is predicted in SMO, since both its antiferromagnetic order and the induced ferroelectric order originate from the body-centered manganese atom in its perovskite structure. This is in contrast to the well-studied material Bismuth Ferrite, in which the magnetic and ferroelectric properties originate from the iron and bismuth atoms respectively. In SMO, the strong co-dependence of structural and magnetic properties induces a wide variety of antiferromagnetic orders, ranging from G-type in bulk to C-type and A-type for strained SMO \cite{lee2010epitaxial}, indicative of competing magnetic interactions. Establishing the magnetic order is an essential ingredient to the understanding of the magnetoelectric coupling. Possessing staggered arrangement of magnetic moments in the sublattices that cancels out macroscopically, makes it challenging for common probes to read or manipulate their states in such antiferromagnets. Although scanning probe techniques such as spin polarized scanning tunneling microscopy, magnetic exchange force microscopy and NV centered microscopy are promising \cite{haykal2020antiferromagnetic,hauptmann2017sensing,kosub2017purely}, they prove to be  technically challenging for the study of such orders at low temperatures.

Here, we study the magnetic order in strained SMO thin films, providing new insights into the magnetic domains, combining magnetic measurements using bulk magnetization studies in addition to the surface probing techniques, mediated by the Spin Hall effect, termed Spin Hall magnetoresistance (SMR). SMR has proven to be an effective tool for studying magnetic order, initially demonstrated for ferromagnetic order \cite{althammer2013quantitative,vlietstra2013spin,das2021coexistence}, and thereafter shown to reveal paramagnetic, antiferromagnetic order and even complex magnetic order in spin-spiral and skyrmionic systems \cite{hoogeboom2017negative,feringa2022observation,aqeel2015spin,phanindra2022spin}. The dependence of SMR on surface magnetic moments, rather than the net magnetization in the material, makes it particularly potent for investigating surface antiferromagnetic order.
We show how this surface magnetic order and its spatial variation can be read by designing devices of different sizes, revealing the nature of the underlying magnetic domains. This is accomplished by modifying the magnetic structure, yielding a sinusoidal dependence of the SMR induced resistivity, both in its amplitude and phase, by rotating an external magnetic field.\\

\section{Results and discussion}

\subsection{Strained SrMnO$_{3}$ surface magnetization}

\begin{figure*}[h!]
    \centering
	\includegraphics[width=\linewidth]{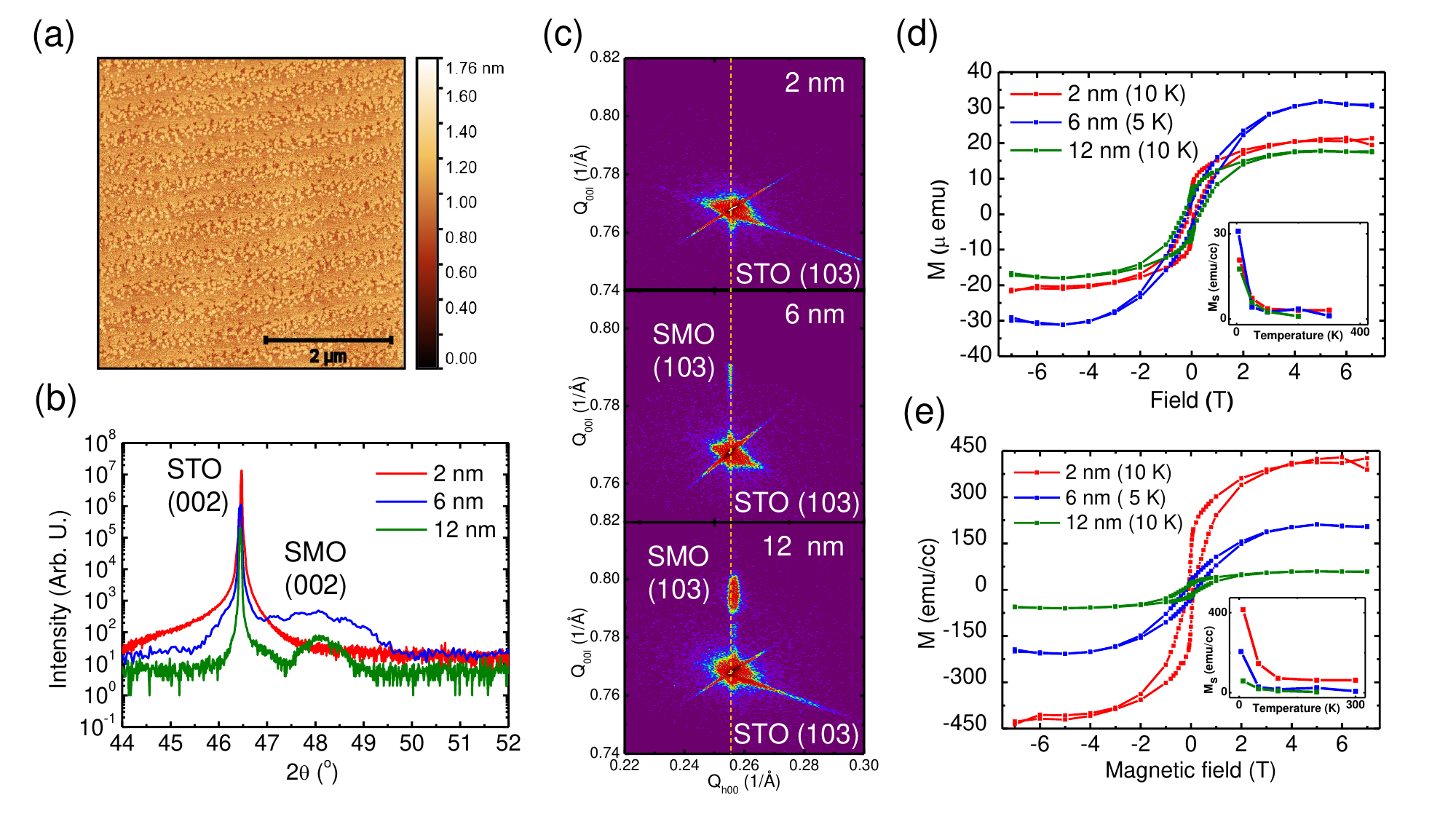}
	\caption{Structural and bulk magnetization results for three SrMnO$_{3}$ films with thicknesses 2 nm, 6 nm and 12 nm. In (a), an atomic force microscopy scan of the 12 nm film shows the surface structure with a roughness of 162 pm indicating an atomically flat surface. In (b) and (c), X-ray diffraction techniques are utilized to confirm the structure of the three films. In (b), 2${\theta}$ scans show the (002) planes and in (c) reciprocal space maps around the SrTiO${_3}$ (103) peak are displayed. Field dependent bulk magnetization measurements are displayed in (d) and (e), where (d) shows the total SMO magnetization which is normalized to the film volume in (e). All M-H hystereses are taken with the field along the [100] crystal direction and are corrected for the linear diamagnetic background from the SrTiO${_3}$ substrate.}
	\label{Fig_1}
\end{figure*}

SrMnO$_{3}$ (SMO) films are grown on SrTiO$_{3}$ substrates using pulsed laser deposition. Three different thicknesses of 2 nm, 6 nm and 12 nm are chosen for the study of bulk magnetic properties in these films. The structural and magnetization measurements are presented in \textbf{Figure \ref{Fig_1}}. The atomic force micrograph displays the surface structure of the 12 nm thick film displaying a smooth film with a root mean square roughness of 162 nm. A four-axis cradle PANalytical x-ray diffractometer (Cu k$\mathrm{\alpha}$ radiation, $\lambda$ = 1.54 \AA) is utilized to establish the structural properties of the films by 2$\mathrm{\theta}$ scans and reciprocal space mapping. No x-ray diffraction peak was observed for the 2 nm thin film.  The Q$_{h00}$ peak positions for the 6 nm and 12 nm films in the reciprocal space maps are situated on the substrate peak as indicated by the dotted line verifying epitaxial growth. From the bulk lattice parameters of STO (3.905 \AA) and SMO (3.805 \AA) \cite{chmaissem2001relationship}, an in-plane tensile strain of 2.6\% is determined. A -0.6\% out-of-plane strain is extracted from the 2$\theta$ scans. This contraction is relatively small with respect to the in-plane strain, implying that strain relaxation mechanisms play a role. For tensile strained SMO in particular, oxygen vacancy formation is known to be an important factor. Although the critical thickness for strain relaxation of SMO films on STO substrates have been determined to be approximately 20 nm \cite{Langenberg2021}, oxygen vacancies effectively relax strain by modifying the valence state of the manganese (Mn) cations at these thicknesses as well. The ionic radius of Mn$^{4+}$ ions is smaller than the oxygen vacancy mediated Mn$^{3+}$ valence state which accommodates for the volume expansion in tensile strained films \cite{Langenberg2021,marthinsen2016coupling,Agrawal2016}.\\

The bulk magnetization of the strained SMO films are studied using a Quantum Design superconducting quantum interference device (SQUID) magnetometer using in-plane magnetic fields up to 7 T at a temperature of 10 K (5 K) for the 2 nm and 12 nm (6 nm) films, shown in figure \ref{Fig_1}d-e. Evidently, all films display a hysteresis curve indicative of a net magnetization originating from ferro-, or ferrimagnetic order. Upon normalizing the total measured magnetization to the respective film volume, figure \ref{Fig_1}e, it is noticeable that the saturation magnetization is strongly dependent on the thickness such that the net magnetization is considerably larger for thinner films. Surprisingly, the curves for the 2 nm and 12 nm films are strikingly similar. The hysteresis curve for the 6 nm film is also comparable, however it displays a larger saturation magnetization (M$_s$) and field which we attribute to the lower measurement temperature of 5 K. The hystereses in figure \ref{Fig_1}d display independence on the film thickness suggesting that the magnetization arises from the film surface or at the interface. We note that any linear field dependence in the bulk magnetization measurements are subtracted, effectively masking any contribution from canted antiferromagnetic moments. The temperature dependent M$_s$ is displayed in the inset.\\
From the structural data, we infer the presence of strain relaxation, most probably due to oxygen defects. Oxygen deficiency drastically affects the magnetic order in strained SMO films, possibly resulting in ferrimagnetic surface magnetization \cite{Kaviani2022,das2021coexistence}. Two reasons are put forth to explain the ferromagnetic hysteresis loops. Firstly, oxygen vacancies create mixed valency of the Mn cations promoting double exchange interaction, arising from the reduction in kinetic energy by electron hopping. Double exchange favouring ferromagnetic interaction, effectively weakens the superexchange and antiferromagnetic exchange interaction. Secondly, oxygen defects modify locally the structure such that the bond angles between the Mn and oxygen can alter up to 30$^{\circ}$, further weakening the antiferromagnetic interaction in accordance with the Goodenough-Kanamori-Anderson rules \cite{van2022strain}. From this we infer that the ferromagnetic hysteresis loops are well explained by the formation of oxygen vacancies. From first principle calculations, the formation energy of oxygen vacancies is lowest in structural domain walls \cite{becher2015strain} and additionally at the surface \cite{Kaviani2022,van2022strain}. Since the amount of oxygen deficiency induced magnetization is independent of the thickness, we argue that a deficient layer forms directly after growth rather than during growth and is inevitable in uncapped strained SMO films. Since the surface magnetization is similar for the three films, we focus on the 6 nm film for probing the local magnetic structure further.

\subsection{Direct probing of magnetically anisotropic antiferromagnetic domains using spin Hall Magnetoresistance}

Although the bulk magnetization reveals the presence of ferromagnetic interaction, the local surface magnetic order is more effectively studied by utilizing a technique that is sensitive to this. In this work, spin Hall Magnetoresistance (SMR) is chosen, which is a result of the interaction at the interface of a heavy metal, in our case Pt, and a magnetic insulator, SMO. Upon application of a current in Pt, the spin accumulation generated by the spin Hall effect is reflected (absorbed) if the surface magnetic moments are ordered parallel (perpendicular) to the spin accumulation, resulting in a relatively low (high) Pt resistivity. The dependence of the Pt resistivity by manipulation of magnetic moments with an externally applied magnetic field is effectively a probe of the surface magnetic order. Important to note here is that with this technique,  only the local magnetic order directly in proximity of the Pt structure is measured.\\ 

\begin{figure*}[h!]
    \centering
	\includegraphics[width=\linewidth]{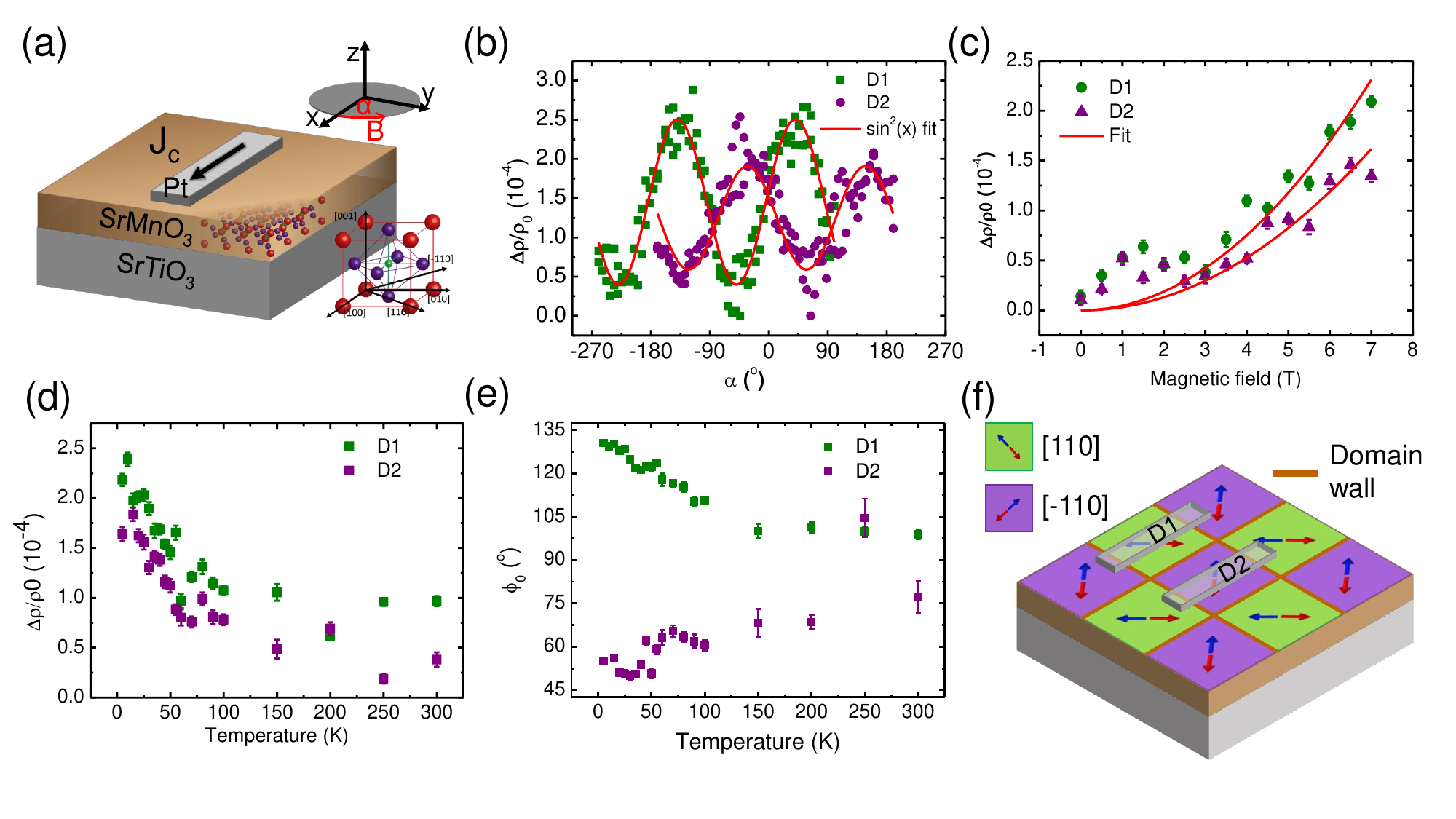}
	\caption{Spin Hall magnetoresistance measurements for two devices on a 6 nm film is displayed, revealing local probing of the antiferromagnetic domains. In (a) the SMR measurement geometry is illustrated. In (b), the angle dependence of both devices at 5 K and an externally applied magnetic field strength of 7 T shows different phase shifts. The quadratic SMR amplitude dependence on magnetic field strength at 5 K is shown in (c) originating from the magnetic domain manipulation. Panels (d) and (e) compare the temperature dependent SMR amplitude and phase respectively for both device. The illustration in (f) visualizes the domain dependent probing of the different devices.}
	\label{Fig_2}
\end{figure*} 

In this work, we explore the dependence of the position and the size of the Pt electrodes in revealing the local magnetic order. \textbf{Figure \ref{Fig_2}} shows the angle dependent magneto resistance response (ADMR) of two small identical Pt devices (7 x 0.5 $\mu$m$^2$) separated by a distance of 1 mm deposited on the 6 nm thick SMO sample. The Pt resistivity is recorded upon applying an AC current of 0.3 mA, as shown in figure \ref{Fig_2}a. The angular dependence of the Pt resistivity when rotating a magnetic field in-plane is displayed in figure \ref{Fig_2}b. Strikingly, the two devices have similar oscillation amplitudes, yet a distinctly different phase. When rotating a field in-plane, the SMR is expected to show a A$\cdot$sin$^2$(x-$\phi_0$) dependence. For ferromagnetic insulators, a phase of 90$^{\circ}$ is expected when a sufficiently large magnetic field aligns the magnetic moments with the field. However, for antiferromagnets, the SMR generally displays a phase of 0$^{\circ}$, since the magnetic moments in the sublattices tend to align perpendicular to the applied field to reduce the Zeeman energy, by canting slightly towards the magnetic field. As is clear from the fit, both devices show a sin$^2$(x) dependence, with a distinct phase of around 135 and 45$^{\circ}$ for device 1 (D1) and device  (D2) respectively. The ADMR is further analyzed by plotting the amplitude as a function of the magnetic field strength and suggests a quadratic dependence \cite{fischer2018spin,geprags2020spin}. The 135$^{\circ}$ phase shift with a quadratic dependence is explained by taking into account preferential magnetic anisotropy axes for oxygen deficient SMO and domain formation \cite{van2022strain}. Here, we note that no spin-flop and spin-flip effects are observed since there is no abrupt change of the SMR signal with field nor a saturation of the SMR amplitude. The [110] magnetic easy axis is stabilized by the symmetry breaking introduced by the oxygen vacancy formation and the resulting Pt resistivity for an in-plane magnetic field rotation is given by

\begin{equation}\label{SMR}
    \rho_{SMR}(\alpha) \propto \rho_{1}\frac{2H^2}{H_{MD}^2}cos^2\left(\frac{1}{4}\pi-\alpha\right) ,
\end{equation}

where $\rho_{1}$ is a SMR coefficient related to the Spin Hall Effect in Pt \cite{vlietstra2013spin}, $H$ is the magnetic field strength, $H_{MD}$ is the monodomainization field and $\alpha$ is the angle of the magnetic field with current direction. The phase shift in D2 located at a different position on the film is then explained by symmetry, since both [110] and [-110] anisotropy axes can be favoured upon formation of an oxygen vacancy. The observation of distinctly different phases in both devices suggest that the small Pt devices directly probe local magnetic moments at the interface with the SMO film, and reveals the direction of their dominant magnetic anisotropy axis, as indicated in Figure \ref{Fig_2}f. This shows that SMR proves to be effective in mapping spatial differences in the local magnetic order in antiferromagnetic films that exhibit magnetic domains. \\
The temperature dependence of the extracted amplitude and phase from the SMR measurements is displayed in Figure \ref{Fig_2}d and \ref{Fig_2}e respectively. With temperature, both devices show a gradual decrease in amplitude which signifies the reduction in magnetic moment with temperature, saturating between 100 K and 150 K, in line with the predicted N\'eel temperature of 121 K, for strained oxygen deficient SMO films \cite{van2022strain}. The phase of both devices is temperature dependent and transitions towards a phase of 90$^{\circ}$ at temperatures above 100 K. We attribute the shift of the phase to temperature dependent magnetocrystalline anisotropy. Above the N\'eel temperature, the phase of both devices stabilizes. Due to the absence of magnetic order, we attribute this temperature independent ADMR signal to Hanle Magnetoresistance, previously observed in Pt nanostructures \cite{velez2016hanle}. From the above discussion, it is evident that the positioning of the Pt devices is crucial to map the spatial differences in the magnetic anisotropy in the different antiferromagnetic domains.\\

\subsection{Device size dependence of Spin Hall Magnetoresistance}

\begin{figure*}[h!]
    \centering
	\includegraphics[width=\linewidth]{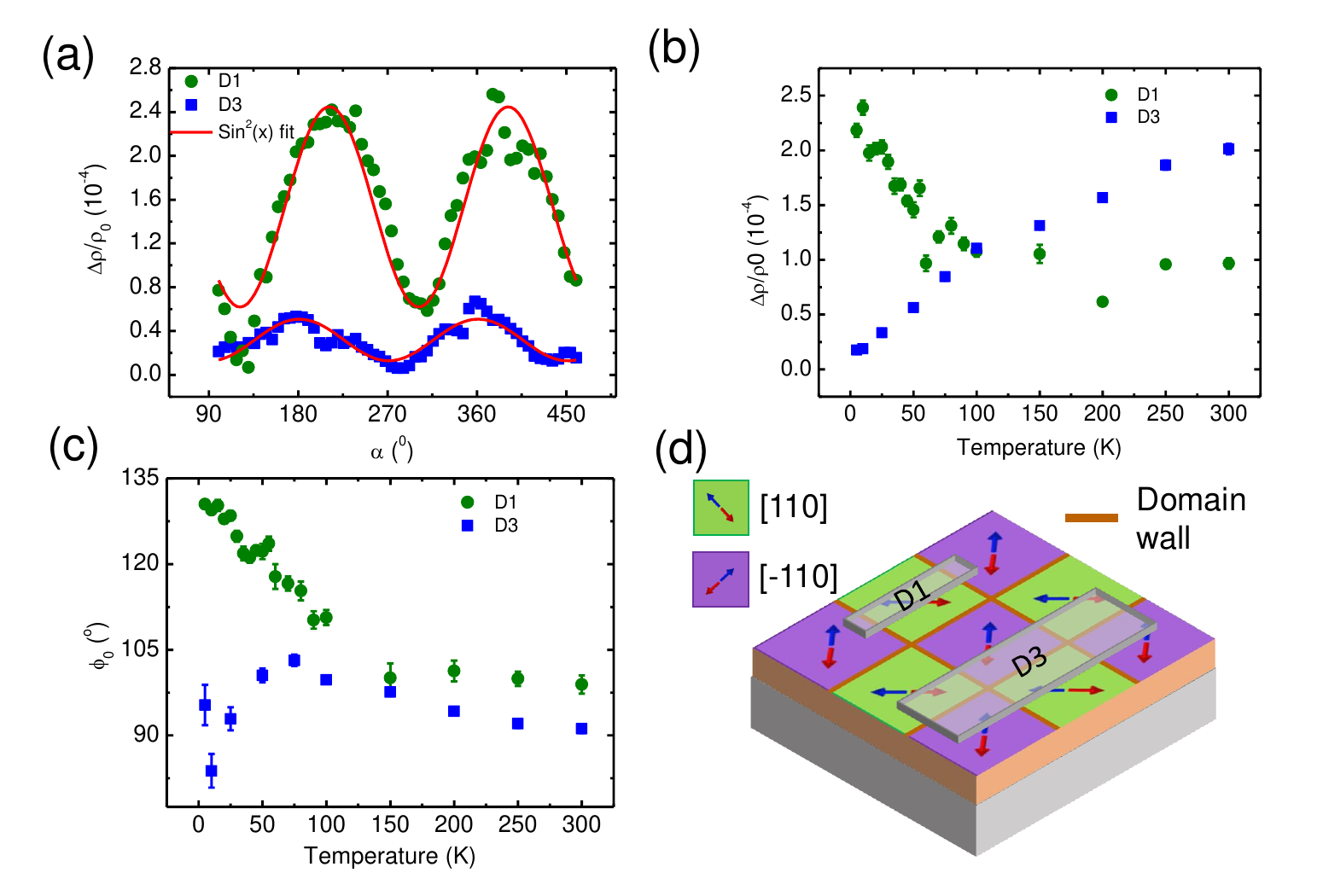}
	\caption{Dependence of SMR on the size of the Pt device on the 6 nm film showing the surface ferromagnetic interaction. In (a), the angular dependence at 5 K and 7 T clearly reveals the difference between the smallest device (7 x 0.5 $\mu$m$^2$) D1 and the largest device (50 x 10 $\mu$m$^2$) D3. The temperature dependent SMR amplitude and phase shift of the two devices are displayed in (b) and (c), taken with a magnetic field strength of 7 T. The illustration in (d) visualizes how the ferromagnetic interaction is dominant for larger devices.}
	\label{Fig_3}
\end{figure*}

Besides the positioning of the Pt devices on the SMO films, the relevance of the Pt device size is investigated, shown in \textbf{Figure \ref{Fig_3}}. In Figure \ref{Fig_3}a-c the SMR response of two devices, the small device (7 x 0.5 $\mu$m$^2$) D1 and a significantly larger device (50 x 10 $\mu$m$^2$) (D3) are studied. Remarkably, the resistivity modulation of the large device, D3, is reduced compared to the response of D1. Secondly, the phase of D3 is extracted to be 90$^{\circ}$, associated with ferromagnetic order, suggesting that this is the dominant contributor for the SMR in the larger device. We propose that the surface magnetization induced by the oxygen vacancies play an important role in the SMR response for large devices. From the ferromagnetic SMR signal, we infer that the contributions from the antiferromagnetic domains with different magnetic anisotropies average out such that the resulting response originates from the remaining net magnetization. The large device encompasses a sufficient amount of differently aligned domains, yielding the non-zero contribution from the aligned magnetic moments.\\

The temperature dependent phase of D3, in Figure \ref{Fig_3}c displays a deviation of around 12$^{\circ}$ with the peak position at 75 K before stabilizing between 100 K and 150 K at 90$^{\circ}$. This phase change is attributed to the disappearance of the double exchange mediated ferromagnetic magnetization (Figure \ref{Fig_1}e inset) at around 50 K. The temperature dependent amplitude in Figure \ref{Fig_3}b for D3 displays an increase of the amplitude with temperature. This increase is attributed to the decrease of the magnetocrystalline anisotropy with temperature, observed also from the gradual phase shift of D1. A decrease of magnetocrystalline anisotropy increases canting of antiferromagnetic moments towards the magnetic field, effectively increasing the net magnetization that contributes to the ferromagnetic SMR response. The discontinuity observed from 100 K agrees well with the N\'eel temperature of 121 K, above which the SMO film is paramagnetic. Considering this, we do not expect the increasing amplitude above 150 K to be related with SMR, and hence the exact nature of the sinusoidal response is unclear. Summarizing, we observe that while the small devices probe the individual antiferromagnetic domains, large devices primarily pick up the net magnetization from double exchange mediated ferromagnetic order and canted antiferromagnetic moments, visualized in Figure \ref{Fig_3}d. From the differences of the measured SMR signals in the small and large device, and the discussions in Section 2.2 we conclude that the domain size is on the order of 3.5 $\mu$m$^2$ from a dominant contribution of a single magnetic anisotropy axis.\\

\section{Conclusion}

In this work, we demonstrate how we decipher the spatial differences in the magnetic order in strained SMO thin films by designing the size and positioning of the Pt devices, from the analysis of the SMR. Surface magnetization contribution in the films can be deduced from the analysis of the observed bulk magnetization studies and arises from the formation of oxygen vacancies due to strain relaxation. Spin Hall magnetoresistance measurements on Pt devices at different locations on the sample reveal local antiferromagnetic domains as proven by the direct observation of both the [110] and [-110] magnetic domains. Interestingly a size dependence of the Pt devices i) reveal a ferromagnetic order for larger device dimension, that arises due to the averaging of the antiferromagnetic order as indicated from the SMR phase and ii) from a complete analysis of the SMR on devices of different dimensions we infer a predominance of antiferromagnetic domain sizes of 3.5 $\mu$m$^2$. \\

\section{Experimental section}

SrMnO$_3$ (SMO) thin films are grown on 5x5 mm SrTiO$_3$ (STO) substrates using pulsed laser deposition. The STO substrates are TiO$_2$ terminated using buffered hydrofluoric acid before an annealing process in oxygen at 960 $^{o}$C. The KrF excimer laser with a wavelength of 248 nm pulses at 1 Hz with a fluence of 2 J/cm${^2}$ in an oxygen pressure of 0.05 mbar and temperature of 800 $^{o}$C. The growth is monitored using in-situ reflective high electron energy diffraction, from which layer-by-layer growth is confirmed by analyzing the diffraction pattern and its intensity oscillations. After growth, the films are annealed for 30 minutes in a 100 mbar oxygen environment at 600 $^{o}$C to optimize the stoichiometry of the magnetic films.\\
The Pt devices are fabricated using electron beam lithography and deposited using DC sputtering. All Pt devices are 7 nm thick and have Ti/Au contacts pads. The Pt resistivity is recorded by applying an AC current with a frequency of 7.777 Hz for all measurements. The current used for the small devices D1 and D2 is 0.3 mA, whereas for all measurements of the larger devices a current of 1 mA is used. \\

\medskip
\textbf{Acknowledgements} \par 
J. J. L. vR acknowledges financial support from Dieptestrategie grant (2019),  Zernike Institute for Advanced Materials. J. J. L. vR and T.B. acknowledges technical support from J. Baas, J. G. Holstein, H. H. de Vries, F. van der Velde and A. Joshua and acknowledges A. Das, S. Chen, A. S. Goossens and A. Jaman. This work was realized using NanoLab-NL facilities. 

\medskip

\bibliographystyle{MSP}
\bibliography{Bibliography}
\newpage

\end{document}